\documentclass[reprint,aps,prb,showpacs,superscriptaddress]{revtex4-1}
\usepackage{mathrsfs,bm,multirow,amsmath,amsfonts,amssymb,array,booktabs,float}
\usepackage{graphicx,times,graphics,color,epsfig}
\renewcommand{\thispagestyle}[1]{}

\begin{document}

\title{Spin-polarized quantum transport properties through flexible phosphorene}
\author{Mingyan Chen}
\affiliation{Department of Physics, Shanghai Normal University, 100 Guilin Road, Shanghai 200232, China}
\affiliation{Department of Physics and the Center of Theoretical and Computational Physics, The University of Hong Kong, Pokfulam Road, Hong Kong SAR, China}
\author{Zhizhou Yu}
\affiliation{Department of Physics and the Center of Theoretical and Computational Physics, The University of Hong Kong, Pokfulam Road, Hong Kong SAR, China}
\affiliation{The University of Hong Kong Shenzhen Institute of Research and Innovation, Shenzhen, Guangdong 518057, China}
\author{Yiqun Xie}
\email{yqxie@shnu.edu.cn}
\affiliation{Department of Physics, Shanghai Normal University, 100 Guilin Road, Shanghai 200232, China}
\affiliation{Department of Physics and the Center of Theoretical and Computational Physics, The University of Hong Kong, Pokfulam Road, Hong Kong SAR, China}
\author{Yin Wang}
\email{yinwang@hku.hk}
\affiliation{Department of Physics and the Center of Theoretical and Computational Physics, The University of Hong Kong, Pokfulam Road, Hong Kong SAR, China}
\affiliation{The University of Hong Kong Shenzhen Institute of Research and Innovation, Shenzhen, Guangdong 518057, China}

\date{\today }

\begin{abstract}
We report a first-principles study on the tunnel magnetoresistance (TMR) and spin-injection efficiency (SIE) through phosphorene with nickel electrodes under the mechanical tension and bending on the phosphorene region. Both the TMR and SIE are largely improved under these mechanical deformations. For the uniaxial tension ($\varepsilon_y$) varying from 0 to 15\% applied along the armchair transport ({\it y}-)direction of the phosphorene, the TMR ratio is enhanced with a maximum of 107\% at the $\varepsilon_y=10\%$, while the SIE increases monotonously from 8\% up to 43\% with the increasing of the strain. Under the out-of-plane bending, the TMR overall increases from 7\% to 50\% within the bending ratio of 0-3.9\%, and meanwhile the SIE is largely improved to around 70\%, as compared to that (30\%) of the flat phosphorene. Such behaviors of the TMR and SIE are mainly affected by the transmission of spin-up electrons in the parallel configuration, which is highly depended on the applied mechanical tension and bending. Our results indicate that the phosphorene based tunnel junctions have promising applications in flexible electronics.

\end{abstract}

%\pacs{71.35.-y, 71.10.Pm, 71.23.An}
\maketitle

Two dimensional (2D) materials\cite{Geim, Reich} have attracted great interests in recent years for their potential applications in various fields, such as logic circuits, solar cells, and photo-detection due to their outstanding electronic and mechanical properties. Most interestingly, the 2D materials is rather desirable for the flexible and wearable consumer electronics because of their out-of-plane flexibility, along with the diverse electronic properties.\cite{Zhu, Chen, Rodin}

Over the past decade, graphene has been the foremost 2D material investigated for flexible nanoelectronics with substantial achievements in large-scale synthesis, device mobility, strain tolerance, and mechanical robustness.\cite{Ko, Rah} However, the lack of a considerable bandgap makes the graphene-based transistors uncontrollable by a gate voltage, being a limitation in applications. Recently, transitional metal dichalcogenides (TMDs) such as MoS$_2$ and WSe$_2$ have emerged as suitable layered semiconductors that offer a sizable bandgap attractive for low-power electronics. The flexible TMDs device has been demonstrated with impressive characteristics,\cite{Das} such as a 45cm$^2$/(Vs) carrier mobility, 10$^7$ on/off ratio, and low contact resistance of 1.4 k$\Omega$-$\mu$m, which are almost unaltered by the in-plane mechanical strain of up to 2\%.

Phosphorene\cite{Li, Chen, Wei, Peng, Wang, Xie, Gong, Zhu, jiwei, Hong, Liu, Fei, Jiang} is a new  member of the 2D material family. It has a higher carrier mobility than the TMDs and a direct band gap determined by the thickness, and shows the anisotropic electronic,\cite{Liu} optical\cite{Hong} and mechanical properties\cite{Jiang} due to its puckered structure, which is very attractive in thin-film electronics. Molecular dynamics simulations have predicted that the phosphorene has an outstanding out-of plane flexibility, which enables the scrolling with a large curvature along the armchair and zigzag directions under mechanical bending while the electronic properties are not severely influenced.\cite{bending} Ref.~\onlinecite{Wei} has also demonstrated that the phosphorene can sustain tensile strain up to 27\% and 30\% in the zigzag and armchair directions, respectively. These studies suggest that the phosphorene is able to maintain a robust electronic and mechanical stability against mechanical tension and bending within a certain extent. Furthermore, a recent experiment has shown that a flexible phosphorene field-effect transistors (FET) can hold a stable electronic properties including mobilities and a fast on/off ratio against the mechanical bending and tension up to 2\%.\cite{Zhu} Nevertheless, the behavior of the spin-polarized electronic properties under mechanical deformations was not examined for this phosphorene-based FET because of its nonmagnetic Ti/Au electrodes. In fact, the study on the spin-polarized transport properties of the phosphorene-based spintronic devices has so far been very limited both in theory\cite{Chen} and experiments. Until recently, the phosphorene-based FET composed of the ferromagnetic TiO$_2$/Co contacts has been fabricated with a predicted tunnel magnetoresistance (TMR) over 10\% based on a modeling calculation,\cite{Ka} while the influence of the mechanical deformations on the spin-polarized transport properties has not been studied yet.

\begin{figure}[tbp]
\includegraphics[width=\columnwidth]{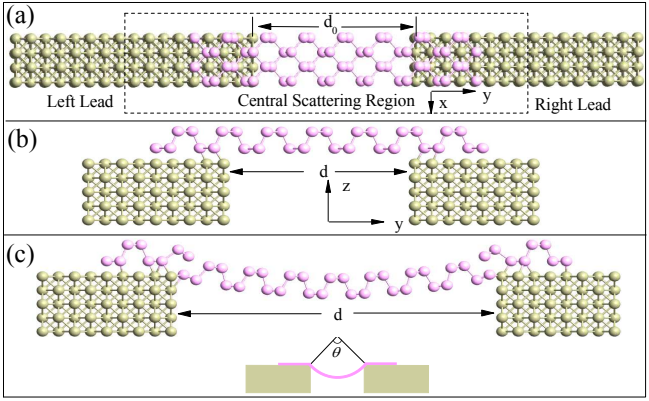}
\caption{(a) The top view of the Ni(100)/phosphorene/Ni(100) MTJ without mechanical strain, (b) the side view of the MTJ with 15\% tensile strain, and (c) the side view of the MTJ with a bending ratio of 3.9\% corresponding to a central angle $\theta$ of 60$^\circ$, as indicated in the inset below. The notations $d$ and $d_0$ denote the distance between the two leads with and without mechanical deformations, respectively. Yellow and pink spheres denote the Ni atoms and  P atoms, respectively.}\label{fig1}
\end{figure}

In this work, to give insights into the behavior of the phosphorene-based flexible spintronic devices, we focus on the influence of the mechanical deformations on the spin-polarized transport properties of the Ni(100)/phosphorene/Ni(100) magnetic tunnel junctions (MTJs), using first-principles quantum transport calculations. In this modeling system [Fig.~\ref{fig1}], the transport is along the armchair direction of the phosphorene, in which the phosphorene can sustain larger mechanical tension and bending than in the zigzag direction as previous studies have proposed.\cite{bending} The variation of the TMR and spin-injection efficiency (SIE) with the mechanical tension and bending are investigated. The TMR and SIE, which mainly affected by the behavior of the transmission of the spin-up electrons in the parallel configuration of the MTJ, are both largely improved compared to those of the original systems without tension and bending. These results indicate that phosphorene-based spintronic devices with Ni contacts has stable performance regarding mechanical tension and bending along the armchair direction, therefore, is very promising to be used to fabricate flexible devices.

The modeling systems of the Ni(100)/phosphorene/Ni(100) are shown in Fig.~\ref{fig1}. Fig.~\ref{fig1}(a) shows the top view of the original MTJ without any strain and bending, and Fig.~\ref{fig1} (b) gives the side view of the MTJ where the mechanical tension is applied along the transport (armchair, $y$) direction. The lattice constants are $a_x=3.524$\AA\ and $a_y=4.58$\AA\ for the phosphorene calculated using the VASP code.\cite{vasp, DFT} The phosphorene is partially overlapped with the Ni(100) electrodes which extend to $y=\pm\infty$. The Ni electrodes are modeled by a five-layer Ni(100) slabs. To build a periodic structure the phosphorene is uniformly stretched by about 6\% along the $x$ direction to match the Ni lattice. The distance between the phosphorene and the Ni slab surface is 1.95{\AA} obtained by VASP.\cite{vasp, DFT} The tension is applied uniformly on the phosphorene along the $y$ direction by increasing the $y$-distance $d$ between the two electrodes. The applied strain is thus defined as $\varepsilon_y$=$(d-d_0)/d_0$, where $d_0$=19.45{\AA} when no strain is applied [See Fig.~\ref{fig1}(a)]. To investigate the influence of the mechanical bending on the transport properties, we constructed a MTJ with a longer phosphorene monolayer with $d_0$=37.77\AA. When the phosphorene is bent, the distance between the two leads (i.e., $d$) is reduced. Therefore, we can also define the applied bending ratio using the distance between the two electrodes as $\varepsilon_b$=$(d_0-d)/d_0$. Fig.~\ref{fig1} (c) shows the side view of the MTJ with a bending ratio of 3.9\%, corresponding to a central angle $\theta$ of 60$^\circ$. The deformed phosphorene was fully relaxed by VASP with the atoms in the Ni electrodes being fixed. We note that in our VASP calculations, the exchange-correlation energy was treated by the projector augmented wave of the Perdew-Burke-Ernzerhof\cite{PBE} with an energy cutoff of 500 eV. The Brillouin zone was sampled with a 8$\times$1$\times$1 mesh of the Monkhorst-Pack k-points.\cite{Mo} A vacuum region of 20{\AA} in the z direction is added in the 2D MTJ supercell to isolate any possible spurious interaction between periodical images of the supercell.

\begin{figure}[tbp]
\includegraphics[width=\columnwidth]{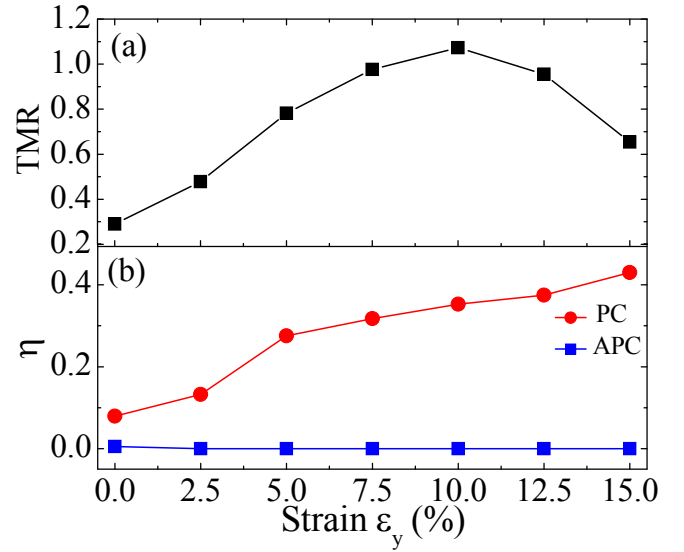}
\caption{The (a) TMR and (b) SIE as a function of the mechanical strain $\varepsilon_y$ for the Ni(100)/phosphorene/Ni(100) MTJ.}\label{fig2}
\end{figure}

Having determined the atomic structures of the 2D MTJ, the spin-polarized quantum transport properties were calculated by the nonequilibrium Green's function formalism combined with the density functional theory (NEGF-DFT) implemented in \texttt{NANODCAL} software package.\cite{NEGFDFT} In the calculations, double-zeta polarized basis is used for Ni atoms and single-zeta polarized basis is used for P atoms; the exchange-correlation were treated at the level of local spin density approximation;\cite{LSDA1, LSDA2} atomic cores are defined by the standard norm conserving nonlocal pseudo potentials;\cite{TM} and 300$\times$1$\times$1 k-points were used to calculate the transmission spectra.

In the following, we analyze two important device merits, namely, the TMR ratio and  SIE. The TMR ratio is  defined as TMR $\equiv (T_{PC}-T_{APC})/T_{APC}$, and the spin-injection efficiency (SIE) is
$\eta\equiv \frac{|T_{\uparrow}-T_{\downarrow}|}{|T_{\uparrow}+T_{\downarrow}|}$. Here, $T_{PC}, T_{APC}$ are the transmission coefficients where the magnetic moments of the two Ni contacts are in parallel configuration (PC) and anti-parallel configuration (APC), respectively; $T_{\uparrow}, T_{\downarrow}$ denote the transmission coefficients by the spin-up and -down channels respectively, which correspond to the minority spin and majority spin for Ni electrodes, and total transmission coefficient is $T_{\uparrow}+T_{\downarrow}$. We note that the transmission values at the Fermi level of the Ni electrodes are used to calculate the TMR and SIE in this work.

\begin{figure}[tbp]
\includegraphics[width=\columnwidth]{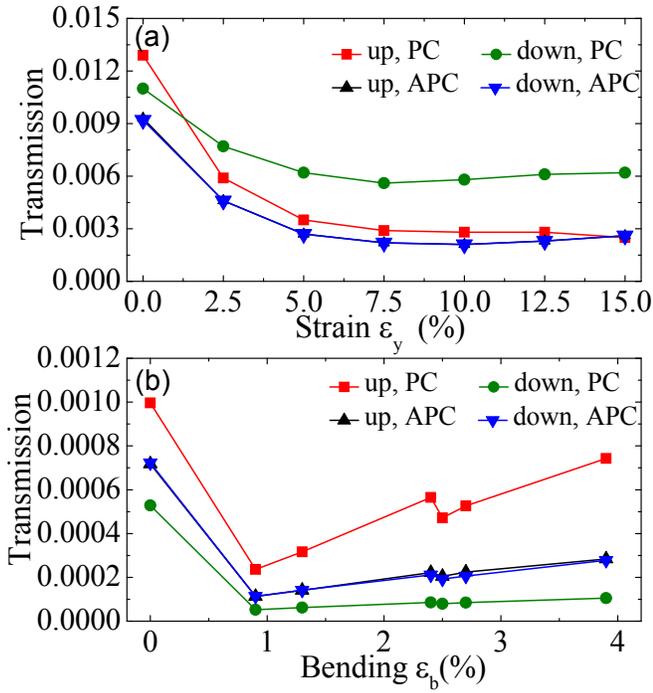}
\caption{The transmission coefficients at the Fermi level for the PC and APC of the Ni(100)/phosphorene/Ni(100) MTJ under the mechanical (a) tension $\varepsilon_y$ and (b) bending $\varepsilon_b$.}\label{fig3}
\end{figure}

\begin{figure}[tbp]
\includegraphics[width=\columnwidth]{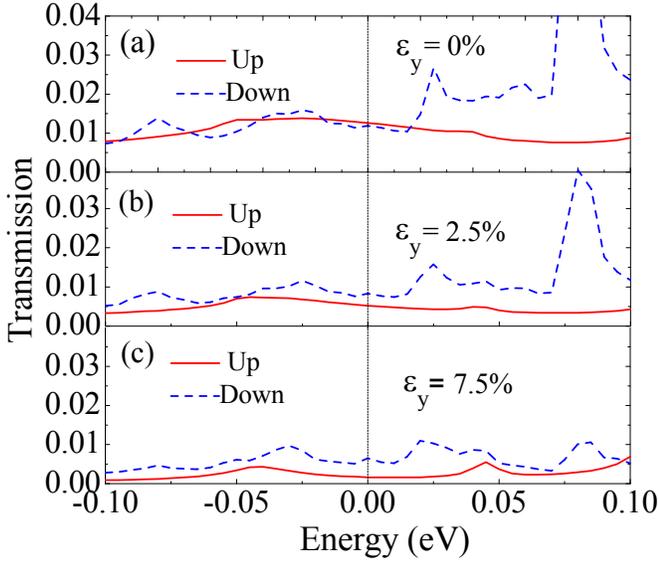}
\caption{Zero bias transmission coefficient versus electron energy in PC of the MTJ with different tensions. The Fermi level is at the energy zero.}\label{fig4}
\end{figure}

We first applied the mechanical tension varying from 0 to 15\% on the phosphorene along the $y$-direction, and calculated the transmission spectrum for different tensions. The TMR and SIE are plotted in Fig.~\ref{fig2}. We observed from Fig.~\ref{fig2} (a) that the TMR increases with the tension $\varepsilon_y$ till 10\%, after which the TMR decreases for the further stretching in the $y$-direction. The maximum of the TMR at $\varepsilon_y$=10\% is about 107\%. Overall, the TMR is enhanced by the mechanical tension with the $\varepsilon_y$ value upto 15\%. Fig.\ref{fig2}(b)
shows that for the PC case (red spheres), the SIE increases monotonously with the $\varepsilon_y$, and reaches about 43\% at $\varepsilon_y$=15\%. For the APC (blue squares), the SIE are almost zero, due to the symmetrical geometry of the MTJ.

Such behaviors of the TMR and SIE under the tension $\varepsilon_y$ can be understood from the change in the transmission at different strains as shown in Fig.~\ref{fig3}. We found that the transmission of the APC (blue down and black up triangles) decreases with the $\varepsilon_y$ from 0 to 10\%, which can account for the increase of TMR in this strain range. After $\varepsilon_y$=10\%, the APC transmission gradually increases, while the spin-up transmission (red squares) of the PC decreases, resulting in a reduction of the TMR. Fig.~\ref{fig3} (a) shows that the PC spin-up transmission  experiences a sharper decline from $\varepsilon_y$=0 to 5\%, as compared to the spin-down component, leading to a sharp decrease of the total transmission ($T_{\uparrow}+T_{\downarrow}$) and thus a faster increase of the SIE in Fig.~\ref{fig2}(b). From $\varepsilon_y$=5\% to 15\%, the spin-up transmission decreases much slower, and accordingly the SIE has a slower increase with the strain. These facts indicate that the spin-up transmission plays a main role in determining the behavior of the TMR and SIE under the mechanical tension. Furthermore, we plot the transmission versus electron energy curves with different tensions in Fig.~\ref{fig4}, from which we can also observe that the transmission value of spin-up electrons is slightly bigger than that of the spin-down electrons in the original length and then more quickly decrease in the whole electron energy range from -0.1 to 0.1 eV with the increasing of the tension. The small spin-up transmission values in a finite energy range indicate a small current value under finite bias voltages.

\begin{figure}[tbp]
\includegraphics[width=\columnwidth]{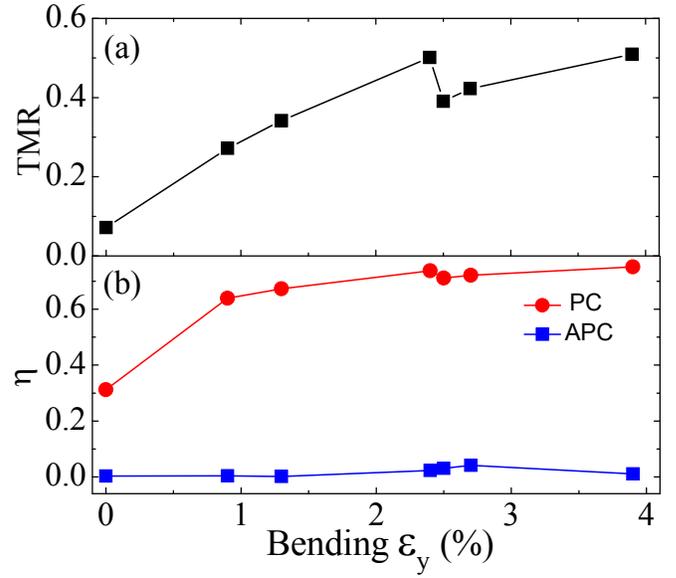}
\caption{The (a) TMR and (b) SIE as a function of the bending ratio $\varepsilon_b$ for the Ni(100)/phosphorene/Ni(100) MTJ.}\label{fig5}
\end{figure}

Next, we examine the influence of the mechanical bending on the TMR and SIE for the Ni(100)/phosphorene/Ni(100) MTJ. The bending is applied in the out-of-plane directions within the range of $\varepsilon_b$=[0, 3.9\%], corresponding to the bending angles from 0 to 60$^\circ$.  The TMR and SIE are given in Figs.~\ref{fig5}(a) and (b), respectively, at different bending ratios.  It shows that both the TMR and SIE are enhanced by the bending. More specifically, overall the TMR and SIE increase monotonously with the bending. The TMR is increased to around 50\% for $\varepsilon_b$ in the range from 2.4\% to 3.9\%, and moreover the SIE of the PC is significantly improved to around 70\%, compared to that (30\%) without bending.

\begin{figure}[tbp]
\includegraphics[width=\columnwidth]{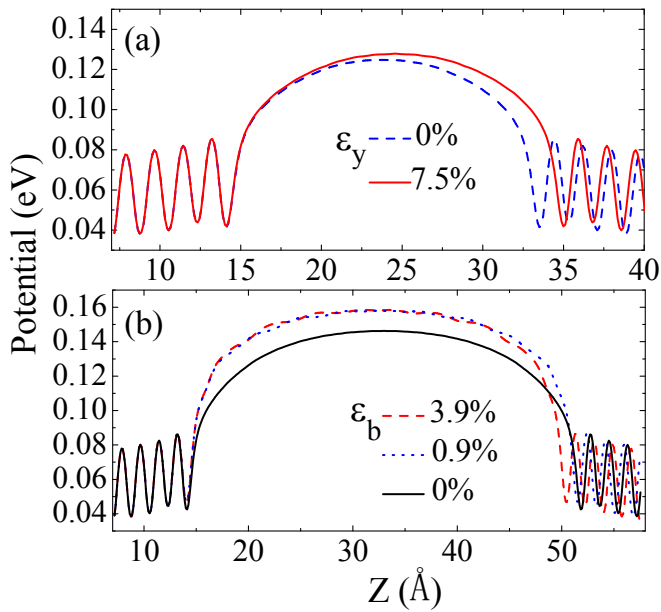}
\caption{The electrostatic potential along the transport direction in the phosphorene region of the Ni/phosphorene/Ni MBJ under various mechanical tension and bending.}\label{fig6}
\end{figure}

The above variation of the TMR and SIE with the bending ratios can also be explained from the transmission shown in Fig.~\ref{fig3}(b). For the small bending ($\varepsilon_b$) from 0 to 0.9\%, the improved TMR and SIE of the PC should be resulted from the decreased spin-up and spin-down transmission, which leads to the reduction of the $T_{PC}$+$T_{APC}$ and $T_{\uparrow}+T_{\downarrow}$. Within the higher bending range of [0.9\%, 3.9\%], by comparing Figs.~\ref{fig5}(a),(b) and Fig.\ref{fig3}(b), we found that the TMR and SIE of the PC have basically the same trend as that of the spin-up transmission for the PC [the red squares in Fig.~\ref{fig3} (b)], while the spin-down transmission (green spheres) is almost invariable with the bending. Therefore, the behaviors of the TMR and SIE induced by the bending are also mainly determined by the spin-up transmission. In addition, the SIE of the APC is still around zero, owing to the symmetrical structure of the MTJ.

After the above microscopic analyses on the origin of the TMR and SIE from equilibrium transmission values, we wish to give a more intuitively explanation from the view point of electron tunneling on the changing of the transmission values. Clearly, the mechanism of the transport in our Ni/phosphorene/Ni MTJ system is electrons tunneling though a finite potential barrier provided by the phosphorene region.\cite{Chen} Fig.~\ref{fig6} plots the electrostatic potential along the transport direction in the phosphorene region of the Ni/phosphorene/Ni MTJ with different mechanical tension and bending, from which different potential barriers for the electrons to tunnel are clearly shown. As in Fig.~\ref{fig6}~(a), after uniaxial tension of 7.5\% is applied the height and width of the potential barrier both increase, resulting in the decreasing of the transmission value [Fig.~\ref{fig3}~(a)]. After out-of-plane bending is applied on the phosphorene, the potential height of the barrier abruptly increases (red dashed and blue dotted lines) compared to that of the original structure (black solid line), leading to the quickly decreasing of the transmission; with further increasing the bending angle, the Ni electrodes approach to each other and the potential barrier width decrease with the potential barrier height keeping nearly unchanged, leading to the gradually increasing of the transmission [Fig.~\ref{fig3}~(b)]. The small kink around 2.5\% of the bending systems may stem from the rearrangements of the atomic bonding during the bending. For atomic level devices, the device performance is strongly dependent on the atomic details, and slightly chemical modifications and changes of the atomic details will lead to fluctuation of the transport properties.\cite{WANGJAP} The kink is normal for our Ni/phosphorene/Ni MTJ which is at atomic length scale. Though the fluctuation exists, the main trends of the results are not affected.

In summary, we have investigated the spin-polarized transport properties of the Ni(100)/phosphorene/Ni(100) MTJ under mechanical tension and bending, using first-principles quantum transport calculations. The uniaxial tension ($\varepsilon_y$) varying from 0 to 15\% is applied along the the armchair direction of the phosphorene. The TMR and SIE of the MTJ are improved by the tension, with the highest TMR of 107\% and SIE of 43\%. Under the out-of-plane bending, the TMR is overall increased monotonously up to about 50\%, and the SIE is substantially improved to around 70\% in a wide bending range of 0.9\% to 3.9\%. These behaviors of the TMR and SIE under the mechanical deformations are mainly influenced by the transmission of the spin-up (minority spin) channels in the parallel configuration. Our results propose that the phosphorene-based MTJs would have stable spin-polarized transport properties against the mechanical deformations, which are highly desirable for the future flexible thin-film electronic devices.

This work is supported by the University Grant Council (Contract No. AoE/P-04/08) of the Government of HKSAR, Special Program for Applied Research on Super Computation of the NSFC-Guangdong Joint Fund, Seed Funding Programme for Basic Research of HKU and NSFC (No. 11404273, 11674231).

\end{document}